\documentclass[showpacs,preprintnumbers,amsmath,amssymb]{elsart}
\usepackage{rotating}
\usepackage{graphicx}
\usepackage{color}
\usepackage{bm}
\usepackage{epsfig}
\begin{document}
\bibliographystyle{try}
\newcommand{\er}{$\pm$}
\newcommand{\be}{\begin{eqnarray}}
\newcommand{\ee}{\end{eqnarray}}
\newcommand{\widt}{\rm\Gamma_{tot}}
\newcommand{\wadd}{$\rm\Gamma_{miss}$}
\newcommand{\gpiN}{$\rm\Gamma_{\pi N}$}
\newcommand{\getN}{$\rm\Gamma_{\eta N}$}
\newcommand{\gkla}{$\rm\Gamma_{K \Lambda}$}
\newcommand{\gksi}{$\rm\Gamma_{K \Sigma}$}
\newcommand{\gNpi}{$\rm\Gamma_{P_{11} \pi}$}
\newcommand{\gDpf}{$\rm\Gamma_{\Delta\pi(L\!<\!J)}$}
\newcommand{\gDps}{$\rm\Gamma_{\Delta\pi(L\!>\!J)}$}
\newcommand{\sqgDpf}{$\rm\sqrt\Gamma_{\Delta\pi(L\!<\!J)}$}
\newcommand{\sqgDps}{$\rm\sqrt\Gamma_{\Delta\pi(L\!>\!J)}$}
\newcommand{\gnsi}{$\rm N\sigma$}
\newcommand{\gNpf}{$\rm\Gamma_{D_{13}\pi}$}
\newcommand{\gNps}{$\Gamma_{D_{13}\pi(L\!>\!J)}$}
\newcommand{\roper}{$N(1440)P_{11}$}
\newcommand{\srma}{$N(1535)S_{11}$}
\newcommand{\trma}{$N(1520)D_{13}$}
\newcommand{\srmb}{$N(1650)S_{11}$}
\newcommand{\trmb}{$N(1700)D_{13}$}
\newcommand{\trmc}{$N(1875)D_{13}$}
\newcommand{\trmd}{$N(2170)D_{13}$}
\newcommand{\fvma}{$N(1675)D_{15}$}
\newcommand{\fvmb}{$N(2070)D_{15}$}
\newcommand{\fvpa}{$N(1680)F_{15}$}
\newcommand{\srpb}{$N(1710)P_{11}$}
\newcommand{\trpa}{$N(1720)P_{13}$}
\newcommand{\trpb}{$N(2200)P_{13}$}
\newcommand{\trpd}{$N(2170)D_{13}$}
\newcommand{\dtpa}{$\Delta(1232)P_{33}$}
\newcommand{\doma}{$\Delta(1620)S_{31}$}
\newcommand{\dtma}{$\Delta(1700)D_{33}$}
\newcommand{\dtmb}{$\Delta(1940)D_{33}$}
\newcommand{\rthe}{A^{1/2}/A^{3/2}}
\newcommand{\amoh}{$A^{1/2}$}
\newcommand{\amth}{$A^{3/2}$}
\newcommand{\broh}{\Gamma^{1/2}_{\gamma p}/\widt}
\newcommand{\brth}{\Gamma^{3/2}_{\gamma p}/\widt}
\newcommand{\btot}{$\Gamma(\gamma p)$}
\newcommand{\Dpi}{\Delta\pi}
\newcommand{\KL}{\Lambda K^+}
\newcommand{\KS}{\Sigma^0 K^+}

\newcounter{univ_counter}
\setcounter{univ_counter} {0}
\addtocounter{univ_counter} {1}
\edef\HISKP{$^{\arabic{univ_counter}}$ }
\addtocounter{univ_counter}{1}
\edef\GATCHINA{$^{\arabic{univ_counter}}$ }

\begin{frontmatter}

\title{Further evidence for \boldmath $N(1900)P_{13}$\unboldmath\
from photoproduction of hyperons}

\author[HISKP,GATCHINA]{V.A.~Nikonov},
\author[HISKP,GATCHINA]{A.V.~Anisovich},
\author[HISKP]{E.~Klempt},
\author[HISKP,GATCHINA]{A.V.~Sarantsev}, and
\author[HISKP]{U.~Thoma}
\\

\address[HISKP]{Helmholtz-Institut f\"ur Strahlen- und Kernphysik der
Universit\"at Bonn, Germany}
\address[GATCHINA]{Petersburg
Nuclear Physics Institute, Gatchina, Russia}

\date{\today}

%-------------abstract----------------

\begin{abstract}
 We report further evidence for $N(1900)P_{13}$ from an analysis of a
large variety of photo- and pion-induced reactions, in particular
from the new CLAS measurements of double polarization observables
for photoproduction of hyperons. The data are consistent with two
classes of solutions both requiring contributions from
$N(1900)P_{13}$ but giving different $N(1900)P_{13}$ pole positions.
$(M-i\Gamma/2) = (1915\pm50)-i(90\pm25)$\,MeV covers both solutions.
The small elasticity of 10\% or less explains why it was difficult
to observe the state in $\pi N$ elastic scattering.

$N(1900)P_{13}$ is a 2-star resonance which is predicted by
symmetric three-quark models. In diquark-quark models, the existence
of the state is not expected.
 \vspace{5mm}   
%----------end of abstract-------------
\end{abstract}

\vskip 5mm

\end{frontmatter}

The flavour structure of baryons and of their resonances is well
described in quark models which assume that baryons can be build
from three constituent quarks. The spatial and spin-orbital wave
functions can be derived using a confinement potential and some
residual interactions between constituent quarks. The best known
example is the Karl-Isgur model \cite{Isgur:1978wd}, at that time a
breakthrough in the understanding of baryons. Later refinements
differed by the choice of the residual interactions: Capstick and
Isgur continued to use an effective one gluon exchange interaction
\cite{Capstick:bm}, Plessas and his collaborators used exchanges of
Goldstone bosons between the quarks \cite{Glozman:1997ag}, while
L\"oring, Metsch and Petry exploited instanton induced interactions
\cite{Loring:2001kx}. A group theoretical analysis by Bijker,
Iachello and Leviatan gave the same complexity of the spectrum of
baryon resonances \cite{Bijker:1994yr}. Quark models, including a
discussion of different decay modes, were reviewed recently by
Capstick and Roberts  \cite{Capstick:2000qj}.

 A common feature of these models is the large number of predicted
states: the dynamics of three quarks leads to a rich spectrum, much
richer than observed experimentally. The reason for the apparent
absence of many predicted states could be that the dynamics of three
quark interactions is not understood well enough. It is often
assumed for instance that, within the nucleon, two quarks may form a
diquark of defined spin and isospin, and that the diquark is a
`stable' object within the baryon. There is a long discussion on the
nature and relevance of the diquark concept; we quote here a few
recent papers \cite{Anselmino:1992vg,Kirchbach:2001de,Jaffe:2003sg,%
Jaffe:2004ph}. Applied to baryon spectroscopy, the diquark model helps
to solve the problem of the {\it missing baryon resonances}.
Santopinto, e.g., calculated the $N^*$ and $\Delta^*$ excitation
spectrum \cite{Santopinto:2004hw} with the assumption that the baryon
is made up from a point-like diquark and a quark. The results match
data perfectly, provided $N^*$- and $\Delta^*$-resonances are omitted from the
comparison that have an one- or two-star PDG \cite{Yao:2006px} ranking
only.

Of course, there is also the possibility that symmetric quark models
treating all three quarks on the same footing are right, and that
the large number of predicted but unobserved states reflects an
experimental problem. In the region between 1900 and 2000\,MeV,
there are 3 two-star resonances, $N(1900)P_{13}$, $N(2000)F_{15}$,
$N(1990)F_{17}$. According to diquark models, these states should
not exist but they are firmly predicted in symmetric three-quark
models. An independent confirmation of the states is therefore
highly desirable.

For long time, the main source of information on $N^*$ and
$\Delta^*$ resonances was derived from pion nucleon elastic
scattering. If a resonance couples weakly to this channel, it could
thus escape identification. This effect may be the reason for the
non-observation of the {\it missing resonances} or for the weak
evidence with which they are observed. Important information is
hence expected from experiments studying photoproduction of
resonances off nucleons, decaying into complex final states. Such
experiments are being carried out at several places. In this letter
we report on further evidence for the $N(1900)P_{13}$, derived from
photoproduction, in particular from recent CLAS data on the spin
transfer coefficients $C_x$ and $C_z$ from circularly polarized
photons to final-state hyperons in the reaction $\gamma p \to
\Lambda K^+$ and $\Sigma K^+$ \cite{Bradford:2006ba}.

The analysis of photoproduction data is not straightforward. Due to
the spin of the initial particles and of the final-state baryon, an
unambiguous solution cannot be obtained without polarization
observables. Moreover, even in the simplest case of single meson
photoproduction a `complete' experiment from which the full
amplitude can be constructed in an energy independent analysis
requires the measurements of at least 8 observables
\cite{Chiang:1996em}. Not only single polarization observables are
required but also double polarization variables need to be measured.
Photoproduction of hyperons is very well suited to measure double
polarization observables since the self-analyzing decay of the
hyperon provides access to the hyperon `induced' polarization, and
only one further observable needs to be determined, e.g. by using a
polarized photon beam.

Recently, the CLAS collaboration measured the spin transfer
coefficients $C_x$ and $C_z$ from circularly polarized photons to
final-state hyperons in the reactions $\gamma p \to \KL$ and $\gamma p
\to \KS$, in the invariant mass region from threshold to $W=2.454$ GeV
\cite{Bradford:2006ba}. These measurements have yielded the first data
expected from  a series of double polarization photoproduction
experiments which are presently planned and carried out at Bonn, JLab,
and Mainz. Even though the new CLAS data provide an important step into
the direction of a complete experiment, we are still far from being
able to reconstruct fully complex amplitudes in a model independent
way. An alternative approach is therefore to include many reactions in
a coupled channel analysis. This direction is followed by EBAC, the
JLab Excited Baryon Analysis Center \cite{Matsuyama:2006rp}, by the
Giessen group \cite{Shklyar:2005xg} and by the Bonn-Gatchina group
\cite{Anisovich:2004zz,Anisovich:2006bc}.

The main input into the new analysis presented here are the new data
on hyperon photoproduction \cite{Bradford:2006ba} in combination
with the analysis of a large number of other reactions. It will be
shown that the data can be described well under an assumption that a
further baryon resonance exists in the 1800-2000\,MeV mass region
which had not been taken into account in our previous fits
\cite{Anisovich:2005tf,Sarantsev:2005tg}. Identification of the new
state with the $N(1900)P_{13}$ is plausible.

Apart from the data on polarization transfer \cite{Bradford:2006ba}, the
following data sets were included in the analysis: differential cross
sections $\rm\sigma(\gamma p \rightarrow \Lambda K^+)$,
$\rm\sigma(\gamma p \rightarrow \Sigma^0 K^+)$, and $\rm\sigma(\gamma p
\rightarrow \Sigma^+ K^0)$, recoil polarization, and photon beam
asymmetry \cite{Glander:2003jw,McNabb:2003nf,Zegers:2003ux,%
Lawall:2005np,Bradford:2005pt,Lleres:2007tx,Castelijns:2007qt};
photoproduction of $\pi^0$ and $\eta$ with measurements of
differential cross sections, beam and target asymmetries and recoil
polarization from the SAID data base \cite{SAID,SAID1,SAID2,%
Krusche:nv,GRAAL2,Bartholomy:04,Crede:04,Bartalini:2005wx}.
Amplitudes for $\pi N$ elastic scattering from \cite{elast} were
included for the low-spin partial waves. The data include about
16.000 data points on two-body reactions; acceptable fits give a
total $\chi^2$ of less than 20.000. A more detailed description of
the analysis method and comparison of the fit with further data can
be found elsewhere \cite{Anisovich:2007}.

Photoproduction of $2\pi^0$ \cite{Thoma:2007,Sarantsev:2007} off
protons and the recent BNL data on $\pi^- p\to n\pi^0\pi^0$
\cite{Prakhov:2004zv} were also included. These data sets were taken
into account in an event-based likelihood fit; at present this data
is restricted to the low-mass region (M$<$1.8\,GeV). The data define
isobar contributions like $\Delta\pi$ and $N\sigma$
\cite{Thoma:2007} and help to disentangle the properties of the
Roper resonance \cite{Sarantsev:2007} but have little influence on
states in the 2\,GeV region. The reaction $\gamma p\to p\pi^0\eta$
was included as well; it provides access to $\Delta P_{33}$ and
$\Delta D_{33}$ partial waves which make the largest contributions
to the latter reaction \cite{Horn:2007}.

The partial wave analysis presented here is based on
relativistically invariant amplitudes constructed from the
four-momenta of particles involved in the process
\cite{Anisovich:2004zz}. High-spin resonances were described by
relativistic multi-channel Breit-Wigner amplitudes, partial waves
with low total spin ($J<5/2$) were described in the framework of the
K-matrix/P-vector approach \cite{aitch}. The $S_{11}$ wave was
fitted as 2-pole 5-channel K-matrix ($\pi N$, $\eta N$, $K\Lambda$,
$K\Sigma$, $\Delta(1232)\pi$); the $P_{11}$-wave as 3-pole 4-channel
K-matrix ($\pi N$, $\Delta(1232)\pi$, $N\sigma$, $K\Sigma$) and
$D_{33}$ wave as 2-pole 3-channel K-matrix ($\pi N$,
$\Delta(1232)\pi$, $S$ and $D$-waves). The $P_{13}$ partial wave is
described alternatively by a sum of Breit-Wigner amplitudes or by a
3-pole 8-channel K-matrix. Amplitudes for elastic $\pi N$ scattering
for the $S_{11}$, $P_{11}$, $D_{33}$, and $P_{33}$ partial waves are
described using the same K-matrix used for photoproduction. The full
parametrization of the $S_{11}$, $P_{11}$, and $P_{13}$ partial
waves is given in \cite{Anisovich:2007}.

Resonances may make large contributions to one reaction and smaller
contributions to other reactions. This property helps considerably
in the identification of resonances and in the determination of
their properties. Table \ref{old} lists the strongest contributions
in the various reactions which are used in the fits. Further
resonances ($N(1675)D_{15}$, $N(1710)P_{11}$, $N(1875)D_{13}$,
$N(2000)F_{15}$, $N(2170)D_{13}$, $N(2200)P_{13}$, $\Delta
(1620)S_{31}$, $\Delta (1905)F_{35}$) were required to get a good
description of the data. Although these states do not contribute
strongly to the differential cross sections, they are needed for the
description of the polarization variables. In most cases the
properties of these states are compatible with the PDG listings. A
few additional high-mass resonances were added to describe the
intensity. However, spins, parities, masses and widths remained
uncertain, and we do not discuss them here.

\begin{table}[pt]
\caption{\label{old}The four strongest resonant contributions (in
decreasing importance) to the reactions included in this analysis.
Resonances contributing less than 1\% to a reaction are not listed.
The contributions are determined for the energy range where data
(see text) exist. Note that the ordering of the states is sometimes
not well defined: it is, e.g., different for solution 1 (chosen
here) and solution 2 discussed below. In some reactions, $t$- and
$u$-channel exchanges provide a significant contribution to the
cross section, too. \vspace{2mm}}
\renewcommand{\arraystretch}{1.3}
\begin{center}
\begin{tabular}{lcccc}
\hline\hline
Reaction & \multicolumn{4}{c}{Resonances}\\
$\gamma p\to N\pi$&\hspace{-1mm}$\Delta(1232)P_{33}$\hspace{-1mm}
&\hspace{-1mm}$N(1520)D_{13}$\hspace{-1mm}&\hspace{-1mm}
$N(1680)F_{15}$\hspace{-1mm}&\hspace{-1mm}$N(1535)S_{11}$\hspace{-1mm}\\
$\gamma p\to p\eta$ &\hspace{-1mm}$N(1535)S_{11}$\hspace{-1mm}&
\hspace{-1mm}$N(1720)P_{13}$\hspace{-1mm}&
$N(2070)D_{15}$& \hspace{-1mm}$N(1650)S_{11}$\hspace{-1mm}  \\
$\gamma p\to p\pi^0\pi^0$ &
\hspace{-1mm}$\Delta(1700)D_{33}$\hspace{-1mm}&
\hspace{-1mm}$N(1520)D_{13}$\hspace{-1mm}
&\hspace{-1mm}$N(1680)F_{15}$\hspace{-1mm}\\
$\gamma p\to p\pi^0\eta$
&\hspace{-1mm}$\Delta(1940)D_{33}$\hspace{-1mm} &
$\Delta(1920)P_{33}$ &\hspace{-1mm}$N(2200)P_{13}$\hspace{-1mm}&
\hspace{-1mm}$\Delta(1700)D_{33}$\hspace{-1mm}        \\
$\gamma p\to \Lambda K^+$&
\hspace{-1mm}$S_{11}$-wave \hspace{-1mm} &
\hspace{-1mm}$N(1720)P_{13}$\hspace{-1mm} &
\hspace{-1mm}$N(1900)P_{13}$ \hspace{-1mm}&
\hspace{-1mm}$N(1840)P_{11}$ \hspace{-1mm}\\
$\gamma p\to \Sigma K$&
\hspace{-1mm}$S_{11}$-wave \hspace{-1mm}&
\hspace{-1mm}$N(1900)P_{13}$ \hspace{-1mm}&
\hspace{-1mm}$N(1840)P_{11}$ \hspace{-1mm}& \\
$\pi^-p\to
n\pi^0\pi^0$&\hspace{-1mm}$N(1440)P_{11}$\hspace{-1mm}
&\hspace{-1mm}$N(1520)D_{13}$\hspace{-1mm}
&\hspace{-1mm}$S_{11}$-wave\hspace{-1mm}\\
[+0.5ex]
\hline \hline
\end{tabular}
\end{center}
\renewcommand{\arraystretch}{1.0}
\end{table}

\begin{figure}[pt]
\begin{sideways}
\begin{tabular}{cc}
\vspace{-5mm}\epsfig{file=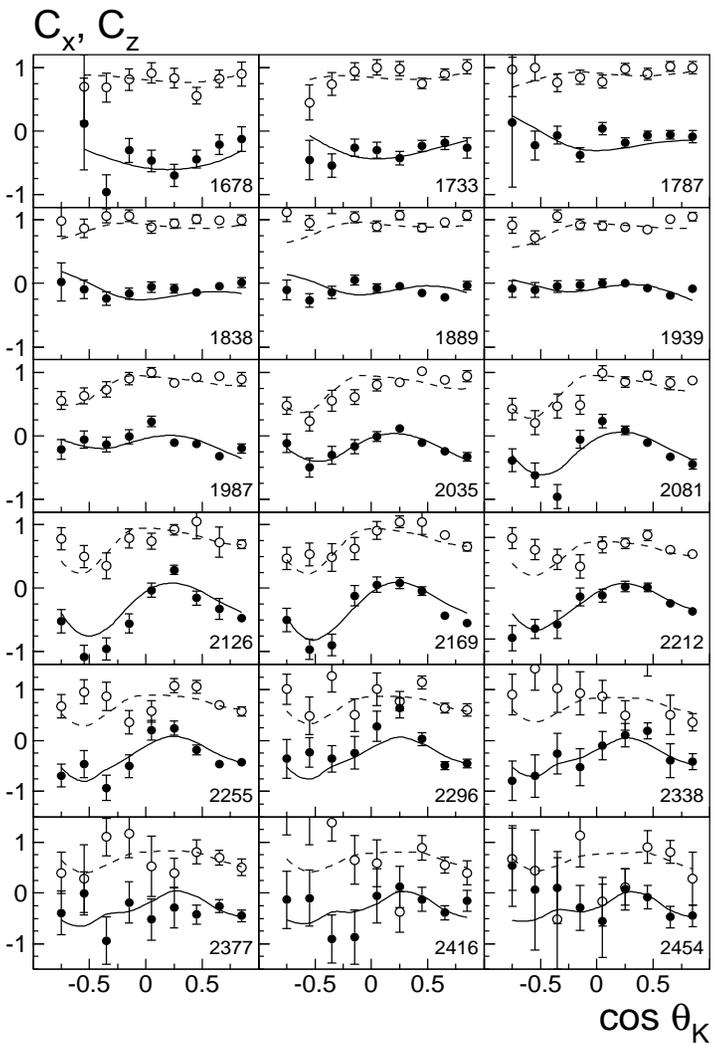,width=0.42\textheight,height=\textwidth}&
\epsfig{file=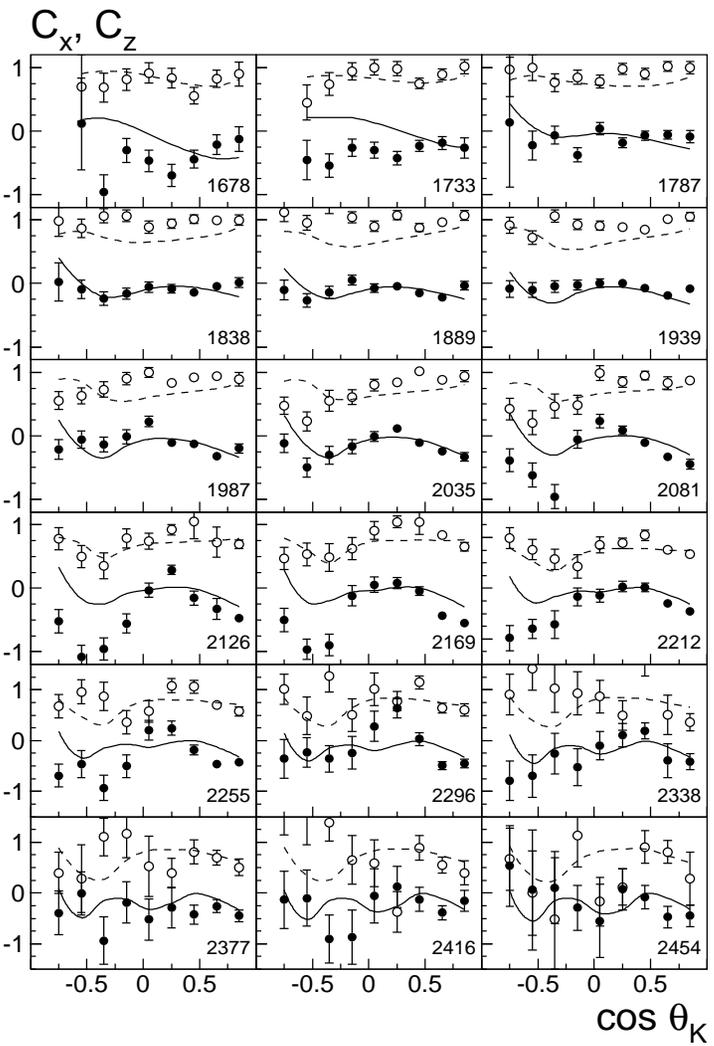,width=0.42\textheight,height=\textwidth}
\end{tabular}
\end{sideways}
\caption{\label{fig:cxcz_klam} Double polarization observables $C_x$
(black circle) and $C_z$ (open circle) for $\rm \gamma p\to\KL$
\cite{Bradford:2006ba}. The solid ($C_x$) and dashed ($C_z$) curves
are our result obtained without (left panel) and with the
$N(1900)P_{13}$ state (right panel) included in the fit.}
\end{figure}

In the first attempt, the data were fitted using one low-mass
Breit-Wigner amplitude to describe the $P_{13}$ wave. A second
$P_{13}$ resonance at M=2200\,MeV was needed to fit the data on
$\gamma p\to p\pi^0\eta$ \cite{Horn:2007}. No good description of
the data was reached. As example, data on $C_x,C_z$ for $\gamma p
\to \KL$ \cite{Bradford:2006ba} are compared to the fit in Fig.
\ref{fig:cxcz_klam}a. Systematic discrepancies are observed
demonstrating the need to introduce further amplitudes. In a second
step we added to our solutions, one by one, Breit-Wigner resonances
with different quantum numbers. The largest improvement was observed
introducing a second $P_{13}$ state. The fit optimized at $1885\pm
25$\,MeV mass and $180\pm 30$\,MeV width, with improvement of
$\chi^2$ for the reactions with two-body final states,
$\Delta\chi^2_{\rm 2b}=1540$ where $\chi^2_{\rm 2b}$ is defined as
the (normalized) sum of the $\chi^2$ contributions of all two-body
reactions, including their weights (see eq. (15) in
\cite{Anisovich:2007}). Adding a $S_{11}$ \{or $D_{15}$\} state
instead, improved the description by 950 \{970\} units. Replacing
the $P_{13}$ by a $P_{11}$ state resulted in a much smaller
improvement, $\Delta\chi^2_{\rm 2b}=205$, probably due to the fact
that the fit included already a $P_{11}$ resonance in this mass
region. A $F_{15}$ state produced a marginal change in $\chi^2_{\rm
2b}$ as well; introducing $F_{17}$ and $G_{17}$ did not improve the
fit. A resonance with $P_{33}$ quantum numbers state provided a
better description of the $\KS$ channel and gave some additional
freedom to the fit of the $\KL$ reaction. However, the change in
$\chi^2_{\rm 2b}$ was again smaller by a factor 2 than the one found
for a $P_{13}$ state.

In a final step, the $P_{13}$ was introduced as 3-pole 8-channel
K-matrix with $\pi N$, $\eta N$, $\Delta(1232)\pi$ ($P$ and
$F$-waves), $N\sigma$, $D_{13}(1520)\pi$ ($S$-wave), $K\Lambda$, and
$K\Sigma$ channels. A satisfactory description of the $C_x$ and
$C_z$ distributions was obtained for both, the $\KL$ (see Fig.
\ref{fig:cxcz_klam}b) and the $\KS$ channel (not shown). The
inclusion of the $N(1900)P_{13}$ resonance was essential to achieve
a good quality of the fit, not only for the new $C_x, C_z$ but also
for other data. The $\chi^2/N_F$ for the differential $\KL$ ($\KS$)
cross section reduced from 2.35 to 2.0 (2.4 to 2.1) when the
$N(1900)P_{13}$ resonance was introduced. Fig.~\ref{fig:cxcz_klam}
shows the best fit without (a) and with (b) $N(1900)P_{13}$
included. When the $P_{13}$-wave was treated as K-matrix,
introduction of a third resonance (representing $N(1900)P_{13}$)
improved $\chi^2$ for $\Lambda K^+$ and $\Sigma K$ data by 1650
units, a significant number. Overall, the fit proved to be
marginally better than the fit using Breit-Wigner amplitudes.

\begin{figure}[pt]
\begin{center}
\epsfig{file=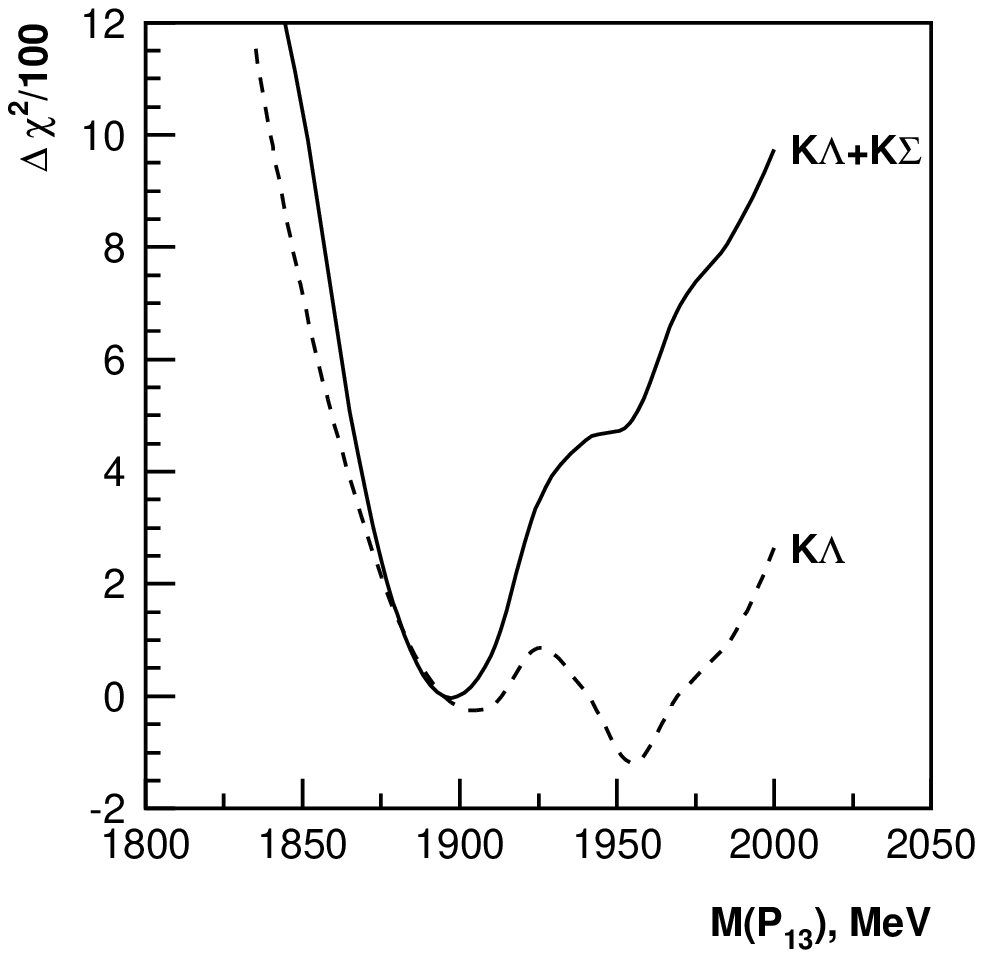,width=0.5\textwidth,height=0.45\textwidth}
\end{center}
\caption{\label{chi2}The change of $\chi^2$ for the fit to
photoproduction of $\Lambda K^+$ and $\Sigma K$ as a function of the
assumed $N(1900)P_{13}$ mass.}
\end{figure}

The $\chi^2$ change as a function of the $N(1900)P_{13}$ mass is
shown in Fig. \ref{chi2}. The $\Lambda K^+$ data exhibit two minima,
corresponding to solution 1 and solution 2, discussed below; the
$\Sigma K$ prefer the lower mass for $N(1900)P_{13}$. Note the
different definitions of the unweighted $\chi^2$ shown in Fig.
\ref{chi2} and the weighted $\chi^2_{\rm 2b}$ used in the fits.

The data set used in this analysis, even though comprising nearly
all available information, is still not yet sufficient to determine
a unique solution. For different start values, the fit can converge
to different minima. As a rule, we accepted all fits which gave a
reasonable description of all data sets and did not show a
significant problem in one of the reactions included. Fits were
rejected, when we found that the trend of the data was inconsistent
with the fit curve even if the increase in $\chi^2$ in some
low-statistics data was counterbalanced by an improved description
of some high-statistics data. When the trend of some data was
inconsistent with the fit curve, we increased the weight of that
data until reasonable consistency was obtained. The variety of
different solutions was used to define the final errors.

All solutions considered from now on include the $P_{13}$ state and
give a reasonable description of all data. However the contributions
of the different isobars to the fitted channels are not uniquely
defined. We observed two classes of solutions which we call the
first and second solution. Both solutions yield a similar overall
$\chi^2_{\rm 2b}$. In the first solution, the pole of the $P_{13}$
partial wave is situated at about 1870\,MeV and provides a
noticeable contribution to the $\KL$ and $\KS$ total cross sections.
It is responsible for the double peak structure in the $\KL$ total
cross section and helps to describe the peak in the $\KS$ total
cross section. In the $\gamma p\to K^0\Sigma^+$ channel, the
contribution of the $P_{13}$ state has a similar strength as the
$N(1840)P_{11}$ state reported in \cite{Sarantsev:2005tg} where the
possible presence of an additional $P_{13}$ state was already
discussed even though it could not yet be identified unambiguously.
In this first solution, the $P_{11}$ pole moved to 1880\,MeV and
became broader. Interference of this pole with the pole at the
region 1700\,MeV generated a comparatively narrow structure in the
$\gamma p\to K^0\Sigma^+$ total cross section.

In the second type of the solutions (the second solution) the $P_{13}$
pole is found at about 1950\,MeV. It provides rather small contributions
to the $\KL$ and $\KS$ total cross sections while the main contribution
to the $\gamma p\to K^0\Sigma^+$ cross section now comes from a
$P_{11}$ state. The new impact of the $P_{13}$ state is an improvement
of the description of double polarization variables due to
interferences. The data are described reasonably well in both
solutions, including those on $C_x$ and $C_z$, see
Fig.~\ref{fig:cxcz_klam}, except perhaps in two slices in the 2.15 GeV
mass region.

A few regions show small but systematic deviations. The first solution
does not describe well the $\KL$ recoil polarization at backward angles
in the 1700\,MeV region. The description can be improved by the
introduction of an additional state in the 1800\,MeV region, with
quantum numbers $\rm P_{33}$, $\rm D_{15}$ or $\rm S_{11}$. In the
latter case, the data might demand a more sophisticated parameterization
of the $S_{11}$ wave by, for example, taking into account the
$\rho(770)N$ threshold. Thus it is not clear if an additional resonance
is really needed. Furthermore, we are not sure that, with the present
quality of the data, these additional states or/and threshold effects
can be identified with reasonable confidence. We therefore decided to
postpone attempts to identify weaker signals until new data are
available. The main result of the present analysis is that a
satisfactory description of the fitted data can be obtained by
introduction of just one new resonance, a relatively narrow $P_{13}$
state at about 1885\,MeV or 1975\,MeV.

The new $P_{13}$ state also improves the description of the $\gamma
p\to K^+\Sigma^0$ reaction, even though its effect is much less
visible here. The double polarization data in this channel were
already described reasonably well in our previous analysis (see the
figures in \cite{Bradford:2006ba}); and a slight readjustment of the
fit parameters gave a good representation of the data. The main
contribution to $\gamma p\to K^+\Sigma^0$ data is now due to
$K$-exchange. In \cite{Sarantsev:2005tg}, the larger contribution
was assigned to $K^*$ exchange. A dominance of $K$ exchange explains
naturally the small $\gamma p\to K^0\Sigma^+$ cross section which is
forbidden for $K$ exchange. The $P_{13}$ partial wave provides a
moderate contribution to the cross section but helps to achieve a
good fit.

\begin{figure}[h!]
\centerline{\epsfig{file=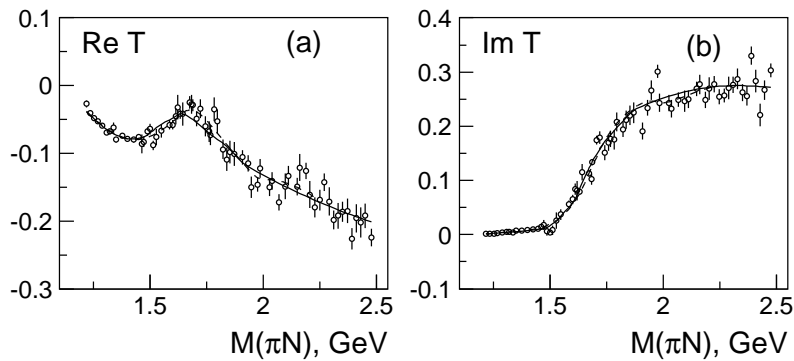,width=0.75\textwidth}}
\caption{Real (a) and imaginary (b) part of the $\pi N$ $P_{13}$
elastic scattering amplitude \cite{elast} and the result of our fit.
Solution 1: solid curve,  solution 2: dashed curve. }
\label{pwa_p13}
 \end{figure}

To check whether elastic data are compatible with the new state, we
introduced it as an additional K-matrix pole and fitted the $\pi
N\to \pi N$ $P_{13}$ partial wave for invariant masses up to 2.4
GeV. A satisfactory description of all fitted observables was
obtained; as example we show the elastic scattering data in Fig.
\ref{pwa_p13}.

Most masses and widths obtained in the fits are compatible with the
numbers given in \cite{Anisovich:2005tf,Sarantsev:2005tg}.  Here we
comment only on the $P_{13}$ partial wave. The parameters of the two
lowest $P_{13}$ poles are given in Table~\ref{Table:p13_all}.  For
both solutions, the first $P_{13}$ state was found to be a rather
broad state. Most previous analyses gave much narrower Breit-Wigner
widths \cite{Yao:2006px}.

However, Manley and Saleski \cite{Manley:1992yb}, the only earlier
analysis which includes $N2\pi$ decays, reported a width of $380\pm
180$\,MeV. The most recent $\pi N\to N\pi$ plus $N\eta$ analysis of
Arndt et al. \cite{elast} gave a pole position at $M=1666, \Gamma =
355$ MeV with no error, not too far from our pole position at
$M=1640\pm 80, \Gamma =480\pm 60$ or (2nd solution) $M=1630\pm 80,
\Gamma =440\pm 60$. Only, the Breit-Wigner masses differ
substantially. Arndt et al. gave $M=1763.8\pm4.6, \Gamma= 210\pm 22$
MeV while we find $1800\pm 100$ ($1780\pm 80$) MeV mass and $700\pm
100$ ($680\pm 80$) MeV width where the numbers correspond to the
first, those in parentheses to the second solution.

The difference in the Breit-Wigner width could indicate a problem.
Attempts to find solutions with a narrower $N(1720)P_{11}$ (with
widths in the 150-250\,MeV range) failed. Yet, Breit-Wigner
parameters are certainly model dependent. It looks strange that in
\cite{elast} the Breit-Wigner width is narrower than the pole width.
We assume that interference between the two $P_{13}$ resonances
leads to an apparent narrowing of the $N(1720)P_{13}$ and
$N(1900)P_{13}$ peaks. If these are fitted using Breit-Wigner
amplitudes, the widths become too narrow. Taking both $P_{13}$
resonances into account in a K-matrix reveals the true
$N(1720)P_{13}$ width. The Breit-Wigner parameters we quote are
derived in a different way: our Breit-Wigner amplitude has exactly
the same pole position as the T-matrix derived from a K-matrix fit.
The state couples strongly to $\Delta(1232)\pi$ and, in the second
solution, also to the $D_{13}(1520)\pi$ channel. The
$D_{13}(1520)\pi$ threshold is close to the resonance mass and
creates a double pole structure. The two poles are hidden under a
Riemann sheet created by a cut at the $D_{13}(1520)\pi$ threshold;
the closest physical region for them is situated above the
$D_{13}(1520)\pi$ threshold. The pole structure renders the
definition of helicity amplitudes and of decay partial widths
complicated; here these quantities are calculated in a procedure
described in \cite{Anisovich:2007} as residues of the poles of the
scattering matrix (T-matrix).

\begin{table}[pt]
\caption{\label{Table:p13_all} Properties of the two lowest $P_{13}$
resonances for both solutions. The masses, widths are given in MeV,
the branching ratios in \% and helicity couplings in 10$^{-3}$
GeV$^{-1/2}$. The helicity couplings and phases were calculated as
residues in the pole position. \vspace{2mm}}
\begin{center}
\renewcommand{\arraystretch}{1.4}
\begin{tabular}{lcccc}
\hline\hline
&\multicolumn{2}{c}{Solution 1}&\multicolumn{2}{c}{Solution 2}\\
\hline
$M_{pole}$               &\hspace{-4mm}$1640\pm 80 $       &\hspace{-3mm}$1870\pm 15$       &\hspace{-3mm}$1630\pm 60 $       &\hspace{-3mm}$1960\pm 15$\\
$\Gamma_{tot}^{pole}$    &\hspace{-4mm}$ 480\pm 80 $       &\hspace{-3mm}$ 170\pm 30$       &\hspace{-3mm}$ 440\pm 60 $       &\hspace{-3mm}$ 195\pm 25$\\
\hline
$A_{1/2}$                &\hspace{-4mm}$ 140\pm 80$        &\hspace{-3mm}$ -(10\pm 15)$       &\hspace{-3mm}$160\pm 40$         &\hspace{-3mm}$-(18\pm 8) $\\
$\varphi_{1/2}$          &\hspace{-4mm}$-(10\pm 15)^\circ$ &\hspace{-3mm}  --               &\hspace{-3mm}$ (10\pm 15)^\circ$ &\hspace{-3mm}$ -(40\pm 15)^\circ $\\
$A_{3/2}$                &\hspace{-4mm}$ 150\pm 80$        &\hspace{-3mm}$-(40\pm 15)$      &\hspace{-3mm}$ 70\pm 30$         &\hspace{-3mm}$-(35\pm 12)$\\
$\varphi_{3/2}$          &\hspace{-4mm}$-(40\pm 30)^\circ$ &\hspace{-3mm}$ (30\pm 25)^\circ$&\hspace{-3mm}$  (0\pm 20)^\circ$ &\hspace{-3mm}$-(40\pm 15)^\circ$\\
\hline
${\rm Br}_{N\pi}$        &\hspace{-4mm}$ 8\pm 4$           &\hspace{-3mm}$ 5\pm  3$         &\hspace{-3mm}$18\pm 5$           &\hspace{-3mm}$ 6\pm 3$\\
${\rm Br}_{N\eta}$       &\hspace{-4mm}$14\pm 4$           &\hspace{-3mm}$20\pm  8$         &\hspace{-3mm}$10\pm 2$           &\hspace{-3mm}$15\pm 3$\\
${\rm Br}_{K\Lambda}$    &\hspace{-4mm}$16\pm 6$           &\hspace{-3mm}$15\pm  5$         &\hspace{-3mm}$ 7\pm 2$           &\hspace{-3mm}$12\pm 3$\\
${\rm Br}_{K\Sigma}$     &\hspace{-4mm}$ <2    $           &\hspace{-3mm}$22\pm  8$         &\hspace{-3mm}$<1     $           &\hspace{-3mm}$ 8\pm 2$\\
${\rm Br}_{\Delta\pi(P)}$&\hspace{-4mm}$54\pm 10$          &         &\hspace{-3mm}$36\pm 6$           &\\
${\rm Br}_{\Delta\pi(F)}$&\hspace{-4mm}$ 2\pm 2$           &         &\hspace{-3mm}$18\pm 5$           &\\
${\rm Br}_{D_{13}\pi}$   &\hspace{-4mm}$2\pm 2$            &         &\hspace{-3mm}$ 5\pm 3$           &\\
${\rm Br}_{N\sigma}$     &\hspace{-4mm}$ 4\pm 2$           &         &\hspace{-3mm}$ 4\pm 2$           &\\
${\rm Br}_{Add}$         &\hspace{-4mm}$<2$ &\hspace{-3mm}$38\pm 12$         &\hspace{-3mm}$ 2\pm 2$&\hspace{-3mm}$60\pm 6$\\
\hline\hline
\end{tabular}
\renewcommand{\arraystretch}{1.0}
\end{center}
\end{table}

The pole of the second $P_{13}$ state is situated in the region
1850-2000\,MeV; it has a smaller coupling to the $\pi N$ channel. In
the first class of solutions, this coupling can be a positive or a
negative value. The helicity couplings are, however, defined under
the assumption that the coupling to the $\pi N$ channel is a
positive number. Thus the sign of the helicity coupling is
ambiguous. In the analysis \cite{Anisovich:2005tf}, only one
$P_{13}$ state below 2.0 GeV was needed to describe the data. This
state was found to be rather broad and to couple to the $\eta n$
channel with branching ratio 8-12\%. The present analysis reproduces
the $P_{13}$ partial wave in the $\gamma p\to \eta N$ reaction even
though the broad structure is produced now due to an interference of
two poles.

In summary, we have analyzed the new CLAS data on spin transfer from
circularly polarized photons to $\Lambda$ and $\Sigma$ hyperons in
the final state. Included in the analysis are other data on photo-
and pion-induced reactions. One additional resonance (compared to
previous fits) is needed to achieve a good description of all data.
Quantum numbers $P_{13}$ are preferred. In spite of the large data
set which includes differential distributions, beam, target and
recoil asymmetries, and some double polarization data, no unique
solution was found. But all solutions require a $P_{13}$ state. The
two classes of solutions from this analysis optimize for masses (and
widths) of 1870 (170) or 1960 (195)\,MeV, respectively. We assign
mass and width of $M=1915\pm 60$\,MeV  and $\Gamma= 180\pm 40$\,MeV
which covers the large majority of all solutions we have obtained.
The elastic widths is about 2-9\%, the branching fraction to
$\Lambda K^+$, 5-15\%.

The Particle Data Group lists two entries for $N(1900)P_{13}$; Manley
and Saleski find mass and width of $1879\pm 17$ ($498\pm78$)\,MeV, the
elastic widths is determined to $0.26\pm 0.06$ \cite{Manley:1992yb}.
Penner and Mosel find $1951\pm 53$ ($622\pm42$)\,MeV and an elastic
width of $0.16\pm 0.02$ \cite{Penner:2002ma,Penner:2002md}. The
$\Lambda K^+$ branching fraction was determined to $2.4\pm 0.3$\%  by
Shklyar and Mosel \cite{Shklyar:2005xg}.

Even though there are considerable inconsistencies between the four
analyses, it seems most likely that the observations are traces of one
resonance. Given its mass and quantum numbers, it can be ascribed to a
quark model state which requires excitation of both oscillators in the
3-body system. The $N(1900)P_{13}$ is unlikely to be explainable in a
picture where a quark is bound by a ``good" diquark.

The work was supported by the DFG within the SFB/TR16 and by a FFE
grant of the Research Center J\"ulich. U. Thoma thanks for an Emmy
Noether grant from the DFG. A.~Sa\-ran\-tsev gratefully acknowledges
the support from Russian Science Support Foundation. This work is also
supported by Russian Foundation for Basic Research 07-02-01196-a and
Russian State Grant Scientific School 5788.2006.2.

\end{document}